\documentclass[aps,pre,twocolumn,groupaddress]{revtex4-1}
\usepackage{color}
\usepackage{graphicx}

\begin{document}
\title{Diffusion in a biased washboard potential revisited}
%
\author{J. Spiechowicz}
\affiliation{Institute of Physics, University of Silesia, 41-500 Chorz{\'o}w, Poland}
\author{J. {\L}uczka}
\affiliation{Institute of Physics, University of Silesia, 41-500 Chorz{\'o}w, Poland}
\email{jerzy.luczka@us.edu.pl}
\begin{abstract} 
The celebrated Sutherland-Einstein relation for systems at thermal equilibrium states that spread of trajectories of Brownian particles is an increasing function of temperature. Here, we scrutinize diffusion of underdamped Brownian motion in a biased periodic potential and analyse regimes in which a diffusion coefficient decreases with increasing temperature within finite temperature window. 
Comprehensive numerical simulations of the corresponding Langevin equation performed with unprecedented resolution allow us to construct phase diagram for the occurrence of the non-monotonic temperature dependence of the diffusion coefficient. We discuss the relation of the latter effect with the phenomenon of giant diffusion.
\end{abstract}
\maketitle

\section{Introduction}
In the Sutherland-Einstein relation \cite{suther,ein}, the diffusion coefficient $D$ of the Brownian particle moving in the viscous medium of temperature $T$ is a linear function of $T$, $D \sim T$, i.e. it increases as temperature grows. This relation is in accordance with our intuition as thermal fluctuations of the medium grow with $T$ and in consequence fluctuations of the Brownian particle position increases as well. The obvious question is whether diffusion can ever slow down with temperature increase? Recent progress in the nonequilibrium statistical physics demonstrates that many phenomena which are forbidden in equilibrium may emerge in nonequilibrium states. Prominent examples include noise-assisted transport \cite{hanggi2009}, negative mobility \cite{slapik2018, slapik2019, slapik2019prappl} and anomalous diffusion \cite{metzler2014}. 

Both experimental data and theoretical studies show that the diffusion coefficient can exhibit non-monotonic dependence on temperature meaning that there is a temperature window in which $D$ decreases when $T$ increases. To the best of our knowledge, the first experimental demonstration was done in 1988 showing that the diffusion of nickel atoms in alloys (chromium-nickel steels) is faster at liquid nitrogen temperature than at room temperature \cite{petrov}. In the subsequent paper \cite{ganshin1999},  the diffusion coefficient in solid $^3\mbox{He} - ^4\mbox{He}$ mixtures at low temperature exhibits a non-monotonic temperature dependence. This  non-monotonicity may be associated with the influence of the non-uniform field of elastic stresses in the crystal due to the difference in the molar volumes of the phases. Similar behaviour has been found in setups including zeolite-guest systems \cite{schuring2002}, cytoplasmic proteins \cite{guo2014} and polymer nanocomposites \cite{tung2016}. It has also been revealed for quantum systems. Spreading of quantum excitations coupled to spatially extended non-linear many-body systems displays non-monotonic temperature dependence which is related to the presence of long-wavelength acoustic modes \cite{iubini2015}. Other examples are diffusive transport in disordered systems in the presence of quantum phonon modes \cite{lee2015} and non-uniform twisted vortex states in rotating superfluids \cite{eltsov2006}. 

In the paper we revisit the problem of diffusion of an inertial Brownian particle in a biased periodic potential. This setup plays a central role in many physical systems \cite{risken}. It models the dynamics of the phase difference across Josephson junction \cite{junction}, rotating dipoles in external fields \cite{Coffey}, superionic conductors \cite{Ful1975}, charge density waves \cite{Gru1981} and cold atoms dwelling in optical lattices \cite{denisov2014}, to mention only a few. 
Various aspects of  diffusion of an underdamped Brownian particle in a washboard potential has been studied \cite{constantini1999, lindenberg2005,kramer,marchenko2014,marchenko2014a,lindner2016,marchenko2017}. 
In some regimes, the system can exhibit two interesting behavior: giant diffusion \cite{constantini1999} and  nonmonotonic temperature dependence of the diffusion coefficient \cite{lindenberg2005}.  Study of these phenomena has been 
continued  \cite{marchenko2014,marchenko2014a,lindner2016,marchenko2017} and explained by bistable velocity dynamics, i.e., in terms of two classes of trajectories for the deterministic counterpart:   locked and running. Analysis has been focused on the phenomenon of giant diffusion to determine for all values of the friction coefficient the range of bias force value for which the diffusion coefficient \emph{apparently} diverges in the zero temperature limit and explained this effect in terms of transition between the locked and running states.  Moreover, some regimes in which \emph{apparently} there is a pronounced maximum of the diffusion coefficient as a function of temperature have been determined. 
 
In contrast, here we concentrate on the effect of non-monotonic temperature dependence of diffusion in this setup to present several complementary results. Most importantly, with the help of numerical simulations of unprecedented resolution we deliver a phase diagram for the occurrence of the non-monotonic temperature dependence of diffusion. We discuss its relation with the corresponding one for the giant diffusion effect and it turns out that it significantly extends the previous predictions.  Moreover, we report a parameter regime for which the velocity dynamics is not bistable, but multistable, thus demonstrating the failure of the bistable velocity dynamics and  the  approximation by a Markovian two-state process \cite{marchenko2014a,lindner2016}.  Last but not least, we highlight the long-lasting transient anomalous diffusive regimes in low temperature region.

The remaining part of the paper is organized as follows. In Sec. II we describe the model and recall quantifiers for the description of diffusion. In Sec. III we present brief history and the state of art of the problem of diffusion in a washboard potential. In Sec. IV we analyse the phenomenon of non-monotonic dependence of the diffusion coefficient on temperature. In particular, we present there a phase diagram for occurrence of this phenomenon. Last but not least, Sec. V contains a summary and conclusions.

\section{Model}
In this paper we reexamine a diffusion process  of a classical Brownian particle of mass $M$, moving in a spatially periodic and \emph{symmetric} potential $U(x) = U(x + L)$ of period $L$ and subjected to a constant biasing force $F$. The dynamics of such a system is described by the following Langevin equation
\begin{equation}
	\label{model}
	M\ddot{x} + \Gamma\dot{x} = -U'(x) + F + \sqrt{2\Gamma k_B T}\,\xi(t), 
\end{equation}
where the dot and the prime denote differentiation with respect to the time $t$ and the particle coordinate $x$, respectively. The parameter $\Gamma$ is the friction coefficient and $k_B$ is the Boltzmann constant. We assume the simplest form of the symmetric potential, namely, 
\begin{equation}
	\label{potential}
	U(x) = -\Delta U \sin{\left(\frac{2\pi}{L}x\right)}
\end{equation}
where $\Delta U$ is half of the barrier height and $L$ is the spatial period. Thermal fluctuations due to the coupling of the particle with the thermal bath of temperature $T$ are modeled by $\delta$-correlated Gaussian white noise $\xi(t)$ of zero mean and unit intensity, i.e.,
\begin{equation}
	\langle \xi(t) \rangle = 0, \quad \langle \xi(t)\xi(s) \rangle = \delta(t-s).
\end{equation}
The noise intensity factor $2\Gamma k_B T$ in Eq. (\ref{model}) follows from the fluctuation-dissipation theorem \cite{kubo1966}.

As the first step of our analysis we transform Eq. (\ref{model}) into its dimensionless form. This can be done  in several ways depending on the choice of the time scale. Here, we define the dimensionless coordinate $\hat x$ and dimensionless time $\hat t$ in the following manner
\begin{equation}
	\label{scaling}
	\hat x = \frac{2\pi}{L} x, \quad \hat t = \frac{t}{\tau_0}, \quad 
	\tau_0 = \frac{L}{2\pi} \sqrt{\frac{M}{\Delta U}}. 
\end{equation}
The characteristic time $\tau_0 = 1/\omega_0$ is the inverse of frequency $\omega_0$ of small oscillations in the well of the potential $U(x)$ defined in Eq. (\ref{potential}). 

Under the above  scaling,  Eq. (\ref{model}) is transformed to the form 
\begin{equation}
	\label{dimless-model}
	\ddot{\hat x} + \gamma \dot{\hat x} = \cos{\hat x} + f + \sqrt{2\gamma \theta}\,\zeta(\hat t),
\end{equation}
where now the dot denotes differentiation with respect to the dimensionless time $\hat t$. We note that the dimensionless mass is $m = 1$ and the remaining rescaled parameters are 
\begin{equation}
	\gamma = \frac{\tau_0}{\tau_1} = \frac{1}{2\pi}\frac{L}{\sqrt{M \Delta U}}\, \Gamma,  \quad f = \frac{1}{2\pi}\frac{L}{\Delta U}\, F. 
\end{equation}
The second characteristic time is $\tau_1= M/\Gamma$ which for a free Brownian particle defines the velocity relaxation time.
The rescaled temperature $\theta$ is given by the ratio of thermal energy $k_{B} T$ to half of the activation energy the particle needs to overcome the original potential barrier $\Delta U$, i.e., 
\begin{equation}
	\theta = \frac{k_B T}{\Delta U}.
\end{equation}
The dimensionless thermal noise $\zeta (\hat t)$ assumes the same statistical properties as $\xi(t)$, namely, it is a Gaussian stochastic process with vanishing mean $\langle \zeta(\hat t) \rangle = 0$ and  the correlation function \mbox{$\langle \zeta(\hat t) \zeta(\hat s) \rangle = \delta(\hat t-\hat s)$}. From now on we will use only the dimensionless variables and shall omit the hat in all quantities appearing in the Langevin equation (\ref{dimless-model}). 
\subsection{Diffusion quantifiers}
The basic quantity characterizing diffusion is the mean square deviation (variance) of the particle coordinate $x(t)$, namely, 
\begin{equation}
	\langle \Delta x^2(t) \rangle = \langle \left[x(t) - \langle x(t) \rangle \right]^2 \rangle = \langle x^2(t) \rangle - \langle x(t) \rangle^2,
\end{equation}
where $\langle \cdot \rangle$ indicates averaging over all thermal noise realizations as well as over initial conditions for the position $x(0)$ and velocity $v(0)=\dot{x}(0)$ of the Brownian particle. The long time evolution of the variance  typically becomes an increasing function of the elapsing time \cite{metzler2014}
\begin{equation} \label{alpha}
	\langle \Delta x^2(t) \rangle \sim t^{\alpha}.
\end{equation}
The exponent $\alpha$ specifies a type of diffusion behaviour. Normal diffusion  is observed for $\alpha = 1$. On the other hand, the case $0 < \alpha < 1$ is termed subdiffusion while for $\alpha > 1$ we classify this behaviour as superdiffusion \cite{metzler2014}. One can define the time-dependent "diffusion coefficient"  $D(t)$ as \cite{spiechowicz2016scirep}
\begin{equation}
	D(t) = \frac{\langle \Delta x^2(t) \rangle}{2t}.
\end{equation}
Note that the case of time-decreasing $D(t)$ corresponds to subdiffusion whereas superdiffusion occurs when $D(t)$ increases. For $D(t) = const.$ we deal with normal diffusion. We stress that only when the exponent $\alpha$ approaches unity we find a properly defined  diffusion coefficient $D$, i.e., \cite{spiechowicz2016scirep}
\begin{equation}
	\label{dc}
	D = \lim_{t \to \infty} D(t) < \infty.
\end{equation}
If the diffusion process is anomalous then $D(t)$ either diverges to infinity (superdiffusion) or converges to zero (subdiffusion) when $t\to\infty$. Therefore the diffusional behaviour of a classical Brownian particle is completely characterized only by both the power exponent $\alpha$ and the diffusion coefficient $D(t)$.

\section{State of the art}
The problem of Brownian motion in a periodic potential has a long history. We refer the interested reader to the well-known Risken  book \cite{risken}. Although at the first glance Eq. (\ref{dimless-model}) may look simple, the Fokker-Planck equation for the particle probability distribution $P(x,v,t)$ corresponding to this equation  is a second order partial differential equation of the parabolic type. Moreover, from Eq. (\ref{dimless-model}) it follows that the parameter space of the model is 
three-dimensional $\{\gamma, f, \theta\}$  which is too large to explore in a systematic and complete way, even with the help of various approximations. Therefore, in particular the issue of diffusion in a periodic potential remains vibrant topic of current research \cite{spiechowicz2019njp, dechant2019, spiechowicz2019chaos, fisher2018, cheng2018, kindermann2017, spiechowicz2017scirep, guerin2017, zhang2017, goychuk2019, spiechowicz2019njp}. Below we briefly review the state of the art of the problem of non-monotonic temperature dependence of diffusion in this setup.
\subsection{Overdamped diffusion}
The comparatively simpler regime that historically was attacked first is an overdamped limit for which inertial effects can be neglected. Such a case is represented by the following  Langevin equation
\begin{equation}
\gamma \dot{x} = \cos{x} + f + \sqrt{2 \gamma \theta}\, \zeta(t).
\end{equation}
In this situation the corresponding Fokker-Planck equation for the particle probability density $P(x, t)$ can be handled analytically and the diffusion coefficient given by Eq. (\ref{dc}) was calculated  exactly in a closed analytical form \cite{lindner2001, reimann2001a, reimann2001b}. It has been shown that the diffusion coefficient in a critically tilted periodic potential may be arbitrarily enhanced as compared to free diffusion $D_0 = \theta/\gamma$. Moreover, already in \cite{lindner2001} the authors reported the non-monotonic temperature dependence of the diffusion coefficient. They argued that this remarkable result relies on the large ratio of relaxation to escape time at the optimal noise intensity.

\subsection{Underdamped diffusion}

For the underdamped dynamics given by Eq. (\ref{dimless-model}) inertial effects play essential role and make the regime much more difficult to analyse. Solutions of the corresponding Fokker-Planck equation for the particle probability distribution 
$P(x, v, t)$ become unattainable. Therefore either approximations or numerical simulations must be applied. Nevertheless the latter may be used only in some particular regions of the parameter space of the model. Amplification of underdamped diffusion was numerically observed due to bistability of the velocity dynamics for the very first time in \cite{constantini1999}. In Ref. \cite{lindenberg2005} it was shown that  maximum of the diffusion coefficient increases rapidly with decreasing friction. Moreover, the authors found that the diffusion coefficient grows with inverse temperature like a power law. This result has been  discussed in \cite{marchenko2014,marchenko2014a}, where the growth of the maximal diffusion coefficient follows rather exponential dependence on inverse temperature. Finally, non-monotonic temperature dependence of the diffusion coefficient for the underdamped dynamics in a tilted periodic potential has been  demonstrated  \cite{lindner2016, marchenko2017}. This salient feature has been recently captured also in systems driven by time-periodic force \cite{spiechowicz2015pre, spiechowicz2016njp, spiechowicz2017chaos, marchenko2018}.

In this study we use extensive numerical simulations of the underdamped Langevin dynamics (\ref{dimless-model}) to construct a phase diagram for occurrence of the non-monotonic temperature dependence of diffusion. In doing so we determine all values of the friction coefficient $\gamma$ as well as the bias $f$ in the numerically accessible parameters range for which the diffusion coefficient display the abnormal dependence on temperature. 
\section{Non-monotonic temperature dependence of underdamped diffusion}
All numerical calculations have been done by the use of a Compute Unified Device Architecture (CUDA) environment implemented on a modern desktop Graphics Processing Unit (GPU). This proceeding allowed for a speedup of factor of the order $10^3$ times as compared to present day Central Processing Unit (CPU) method \cite{spiechowicz2015cpc}. We employed a weak 2nd order predictor-corrector scheme to simulate stochastic dynamics given by Eq. (\ref{dimless-model}) with the time step $h = 10^{-2}$. The initial position $x(0)$ and velocities $v(0)$ were uniformly distributed over the intervals $[0,2\pi]$ and $[-2,2]$, respectively. The quantities characterizing  diffusive behaviour of the system were averaged over the ensemble of $10^5$ trajectories, each starting with different initial condition according to the above distributions.

\subsection{Bistability of the velocity dynamics}
\begin{figure}[t]
	\centering
	\includegraphics[width=0.9\linewidth]{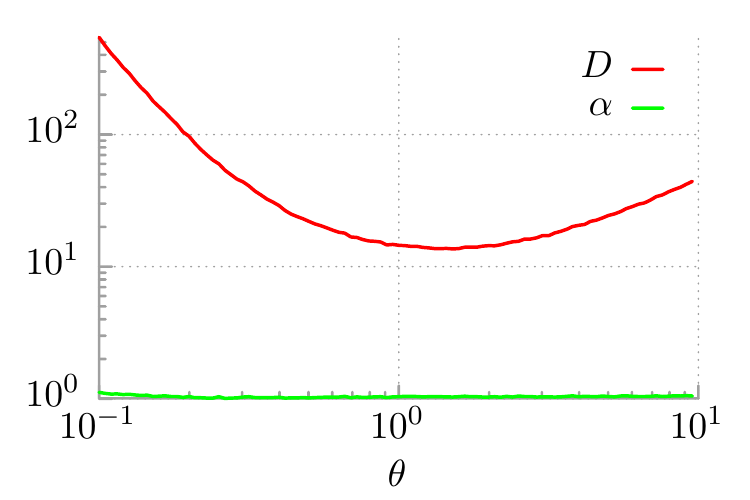}
	\caption{Non-monotonic temperature dependence of underdamped diffusion. The diffusion coefficient $D$ as well as the power exponent $\alpha$ is plotted as a function of the dimensionless temperature $\theta \propto T$ in the parameter regime corresponding to the bistable velocity dynamics. Parameters are: $\gamma = 0.255, f = 0.52$.}
	\label{fig1}
\end{figure}

In Fig. \ref{fig1} we exemplify the non-monotonic temperature dependence of the diffusion coefficient $D$. There, the latter quantifier is plotted versus the dimensionless temperature $\theta \propto T$. The reader can note that already for the whole decade the diffusion coefficient is decreasing with increasing $\theta$. Its minimum is attained for $\theta \approx 1$. Moreover, in the same panel we also depict the asymptotic value of the power exponent $\alpha$ defined in Eq. (\ref{alpha}) to illustrate that indeed the diffusive behaviour is normal in the whole analysed temperature range. The latter fact means that the system has reached its stationary state. One needs to carefully check this for each parameter regime separately to not confuse enhancement of normal diffusion with the occurrence of superdiffusion \cite{spiechowicz2016scirep}. It has been suggested that the mechanism which was responsible for this unconventional non-monotonic diffusive dependence had its roots in the bistability of the thermal noise driven velocity dynamics \cite{marchenko2014a,lindner2016}. In the following we complement the previous analysis  by demonstrating a direct numerical proof for this statement.

\begin{figure}[t]
	\centering
	\includegraphics[width=0.9\linewidth]{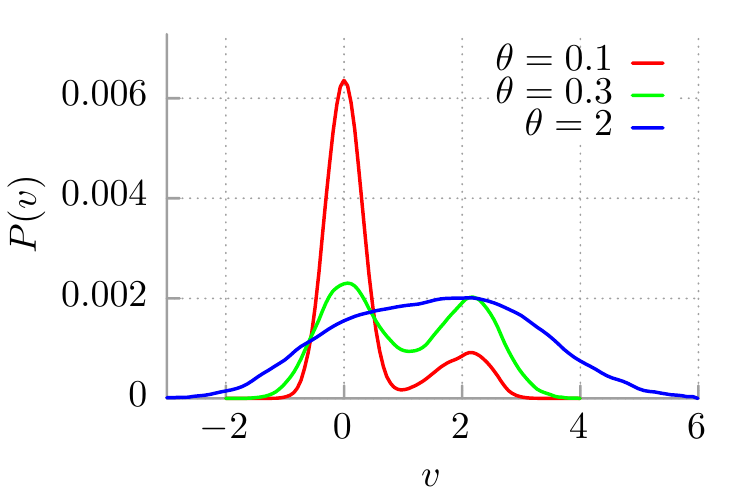}
	\caption{Influence of temperature on bistability and its  disappearance visualized for  the probability distribution $P(v)$ of the particle  velocity in the long time limit, cf. Eq. (\ref{pst}), is depicted in the parameter regime corresponding to Fig. \ref{fig1}, i.e. \mbox{$\gamma = 0.255$ and  $f = 0.52$}. The dimensionless temperature reads $\theta = 0.1$ (red),  $\theta = 0.3$ (green) and $\theta =2$ (blue).}
	\label{fig2}
\end{figure}

First, let us study the corresponding probability distribution for  velocity of the Brownian particle in the long time limit, namely, 
\begin{equation}
	\label{pst}
	P(v) = \lim_{t\to\infty} P(v,t), \quad P(v,t) = \int_{-\infty}^{\infty} dx \,P(x, v, t),
\end{equation}
which is depicted in Fig. \ref{fig2}. Indeed, we note that for the model parameters corresponding to Fig. \ref{fig1} the velocity dynamics is bistable at low to moderate temperature. There are two solutions: the first describes the \emph{locked} one in which the particle is confined in a finite number of potential wells with $v_l \approx  0$ whereas the second is \emph{running} for which the motion is unbounded in space with $v_r \neq 0$. In the presented case the reader can observe a typical finite temperature effect, namely, the broadening of the Gaussian peaks corresponding to each solution. The higher temperature of the system is, the peak corresponding to the locked state decreases and the peak corresponding to the running state increases.  Nevertheless still we are able to  distinguish between the locked solution $v_l = 0$ and the running one $v_r \approx 2$. 
Finally, at high temperature, thermal fluctuations destroy locked states and only running states exist (the blue curve in Fig. \ref{fig2}).  
\begin{figure}[t]
	\centering
	\includegraphics[width=0.9\linewidth]{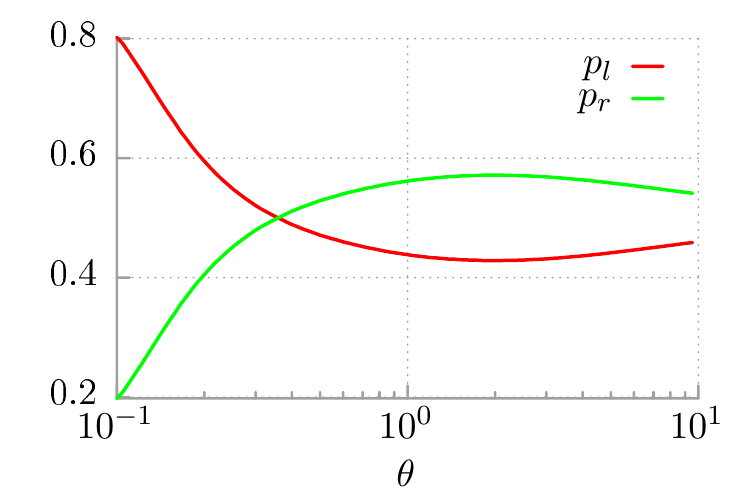}
	\caption{Temperature dependence of the stationary probabilities $p_l$ and $p_r$ for the particle to reside in the locked $v_l = 0$ and running $v_r = 2$ state, respectively. Parameter regime corresponds to Fig. \ref{fig1}, i.e. $\gamma = 0.255$ and $f = 0.52$.}
	\label{fig3}
\end{figure}

In the bistability regime, the above analysis allows to construct a two state Markov process which approximates the Langevin equation (\ref{dimless-model}). Jumps between the locked and running states are driven \emph{solely} by thermal fluctuations. The diffusion coefficient is a measure of a spread of trajectories of the system around its mean path. For our case of the bistable velocity dynamics there are two contributions to it. The first is associated with the spread coming from relative distance between the locked and running trajectories. The second is related to the spread of trajectories following a given velocity solution. The latter is caused solely by thermal fluctuations. The first contribution is much larger than the second, at least for low to moderate temperature. From this reasoning it is clear that the diffusive behaviour of the system in the case of bistable velocity dynamics is significantly impacted by the fraction of locked trajectories. We now consider the stationary probabilities $p_l$ and $p_r$ for the particle to reside in the locked $v_l = 0$ and running $v_r = 2$ states, respectively. In Fig. \ref{fig3} we depict these quantities as a function of temperature $\theta$ of the system in the parameter regime corresponding to Fig. \ref{fig1} for which the diffusive behaviour displays non-monotonic character. We note that the temperature dependence of the probability $p_l$ for the particle to be in the locked state resembles very much behaviour of the diffusion coefficient depicted in Fig. \ref{fig1}. In particular, the overall spread of trajectories is large when there are many locked trajectories. The diffusion coefficient attains its minimum when the fraction of the locked solutions is minimal, c.f. Fig. \ref{fig1}. Further growth of temperature causes an increase of $p_l$ which immediately enlarges the diffusion coefficient. For high enough $\theta$ both stationary probabilities are expected to be equal $p_l = p_r$. However, in such a limit the spread of a group of running trajectories dominates the diffusion coefficient and the latter follows the well known Sutherland-Einstein relation $D \propto \theta$.

\begin{figure}[t]
	\centering
	\includegraphics[width=0.9\linewidth]{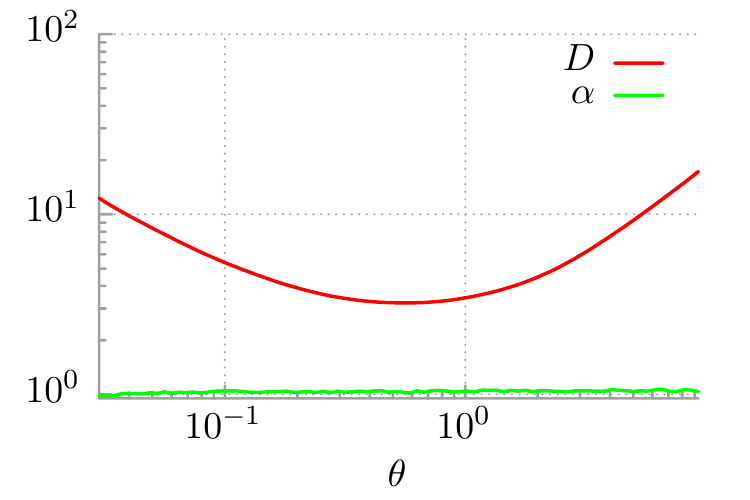}
	\caption{Non-monotonic temperature dependence of underdamped diffusion. The diffusion coefficient $D$ and the power exponent $\alpha$ is shown versus temperature $\theta$ in the parameter regime corresponding to the multistable velocity dynamics. Parameters are $\gamma = 0.6366$, $f = 0.91$.}
	\label{ivan}
\end{figure}
\begin{figure}[t]
	\centering
	\includegraphics[width=0.9\linewidth]{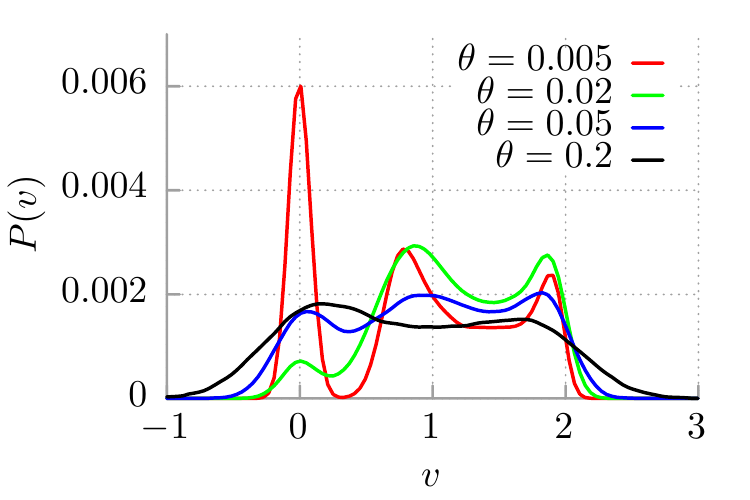}
	\caption{Multistability of the probability distribution for the velocity of the Brownian particle $P(v)$ in the long time limit is presented in the parameter regime corresponding to Fig. \ref{ivan} for different temperature $\theta$.}
	\label{ivan_hist}
\end{figure}
\subsection{Multistability of the velocity dynamics}
In this section we demonstrate that care needs to be taken as one could encounter in the parameter space regimes for which the velocity dynamics can be  multistable, meaning that there exists more than one running solution pointing into the positive direction. Moreover, among these parameter sets one may also discover a non-monotonic temperature dependence of the diffusion coefficient. We exemplify this feature in Fig. \ref{ivan} where we depict the diffusion coefficient $D$ as well as the power exponent $\alpha$ versus temperature $\theta$. In the subsequent panel, i.e. Fig. \ref{ivan_hist}, we present the corresponding probability distribution $P(v)$ for the Brownian particle velocity in the long time limit for different temperature $\theta$. The reader can clearly distinguish two running states transporting the particle into the positive direction which differ by the velocity magnitude. Additionally there is also the locked solution so that the overall velocity distribution is multistable. Moreover, one can note that this behaviour is characteristic for low temperature as for increasing thermal noise intensity the distribution first becomes bimodal and then, for sufficiently high temperature, unimodal. This complexity is characteristic feature of the noisy system as for its deterministic counterpart it is already well known that the velocity dynamics is at most bistable \cite{risken}. It has been overlooked in the previous research, especially in Ref. \cite{lindner2016}, where the authors represented the dynamics as a Markovian two-state process in the velocity space. Clearly, the parameter regime reported in Fig. \ref{ivan_hist} demonstrates the failure of this approximation. In order to understand the mechanism standing behind the non-monotonic temperature dependence of diffusion in this regime most likely one needs to carefully investigate temperature influence on all transitions between the observed three states. We expect that still the pivotal role is played by the jumps between the running solutions and the locked one. However, since such investigation does not lay in the main context of the present paper we leave this problem open for near future research.

\subsection{Transient anomalous diffusion}
\begin{figure}[t]
	\centering
	\includegraphics[width=0.9\linewidth]{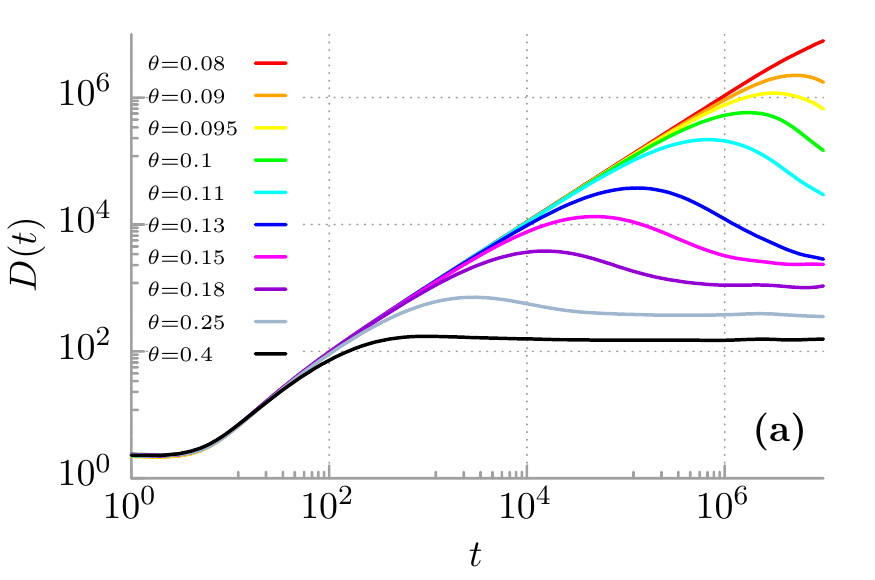}\\
	\includegraphics[width=0.9\linewidth]{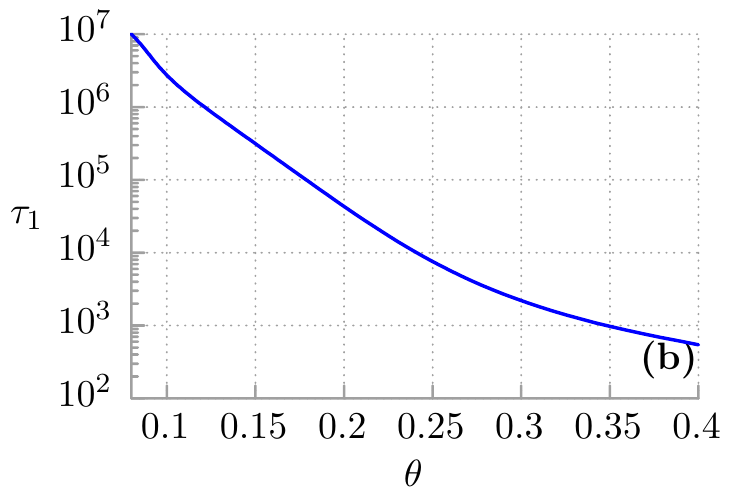}
	\caption{Panel (a): Evolution of the diffusion coefficient $D(t)$ depicted for different values of the dimensionless temperature $\theta \propto T$. Panel (b): Lifetime  $\tau_1$ of the ballistic diffusion versus temperature $\theta$. Other parameters are $\gamma = 0.1$, $f = 0.3$.}
	\label{fig4}
\end{figure}

In Fig. \ref{fig4} (a) we depict the  time dependent diffusion coefficient $D(t)$ for selected values of the dimensionless temperature $\theta$ in another parameter regime for which $\gamma = 0.1$ and $f = 0.3$. The only reason for such a choice of the parameter values is a more pronounced visualization of transient effects of the system dynamics associated with an  investigated non-monotonic temperature dependence of the diffusion coefficient. The latter is  extremely important to understand technical limitations of the numerical study of this system. In panel (a) one can observe two regimes: (i) the first is for high temperatures for which at initial stage superdiffusion (ballistic diffusion) occurs in the time interval $(0,\tau_1)$ where the diffusion coefficient $D(t)$ increases as a function of time and next normal diffusion is approached for $t > \tau_1$. For these regimes $D(t)$ monotonically tends to its time-independent stationary value $D$; (ii) the second is for lower temperatures for which the reader may detect a characteristic {\it non-monotonic} relaxation of the diffusion coefficient $D(t)$ towards its stationary value $D$. Initially $D(t)$  grows with time (ballistic diffusion), then it decreases (subdiffusion) and finally it reaches its steady state $D$ (normal diffusion). The latter relaxation pattern of $D(t)$ is characteristic feature of a non-monotonic temperature dependence of the  diffusion coefficient.

The duration of this transient anomalous diffusive behaviour is extremely sensitive to temperature variation. In particular, as it is shown in panel (b) of the same figure, if temperature decreases the period $\tau_1$ of ballistic diffusion rapidly increases and tends to infinity $\tau_1 \to \infty$ when $\theta \to 0$. Therefore, from the technical point of view, smaller temperatures require exponentially larger simulation times until the diffusion coefficient settles. Transitions between velocity states are driven by thermal noise. Its intensity reads $Q = \gamma \theta$. For smaller intensity jumps between the velocity solutions are less probable. Therefore the above reasoning applies also to the dimensionless friction coefficient $\gamma$, meaning that to be able to reach stationary state in the time span of numerical simulations one has to avoid the limit of small $\theta$ and/or $\gamma$. Another critical point occurs for $f = 1$ for which velocity bistability is no longer observed. The limit $f \to 1$ is also hard to tackle with direct numerical simulations since then the transition rates describing the locked solution become vanishingly small and one again needs exponentially longer time span of the simulation to sample the state space of the system reliably.  
\begin{figure}[t]
	\centering
	\includegraphics[width=1.0\linewidth]{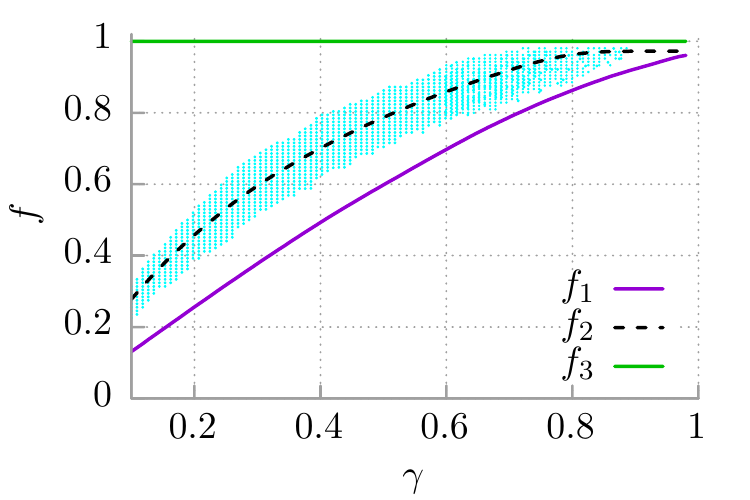}
	\caption{Phase diagram for occurrence of the non-monotonic temperature dependence of diffusion presented in the parameter plane $(\gamma,f)$ together with Risken plot with the three critical forces $f_1, f_2$ and $f_3$.}
	\label{fig5}
\end{figure}

\subsection{Phase diagram of the non-monotonic temperature dependence of diffusion}
We have taken into account the above constraints and constructed the phase diagram of occurrence of the non-monotonic temperature dependence of diffusion in the system described by the Langevin equation (\ref{dimless-model}). This is an essentialy new result which has not been presented in the literature. It was possible solely thanks to the use of our innovative computational method which allowed to explore the parameter space with sufficient resolution \cite{spiechowicz2015cpc}. We performed scans of the following area of the parameter space $\gamma \times f \in [0.1,1] \times [0,1]$ at a resolution 100 points per dimension. For each pair $(\gamma, f)$ we calculated temperature dependence of the diffusion coefficient $D(\theta;\gamma,f)$ in the interval $\theta \in [0.1,10]$. In each case we checked that asymptotically diffusion is normal, i.e. its coefficient is constant $D = const.$ and the power exponent $\alpha = 1$. For the most parameter regimes the span of $10^5$ dimensionless units of time was sufficient to reach the steady state of the system, c.f. Fig. \ref{fig4}. Then we computed the characteristic $\partial{D}/\partial{\theta}$ and verified whether there is a finite interval for which the diffusion coefficient is a decreasing function of temperature, i.e $\partial{D}/\partial{\theta} < 0$. The result is shown in Fig. \ref{fig5}. Each cyan dot represents a pair $(\gamma, f)$ for which the diffusion behaviour of the system displays a non-monotonic dependence on temperature.

According to the state of the art there are two regimes of the latter effect. The first one is associated with the phenomenon of giant diffusion for which the diffusion coefficient $D$ is enhanced and exceeds the bare diffusion coefficient $D_0 = \theta/\gamma$ of a free particle by orders of magnitude if the biasing force is close to a critical value \cite{lindner2001,kramer,reimann2001a, lindner2016}. This amplification is particularly evident for low temperatures $\theta \to 0$ for which the diffusion coefficient \emph{apparently} diverges $D \to \infty$.  This is indeed an example of a non-monotonic temperature dependence. This effect occurs due to  bistability of the velocity dynamics. The latter phenomenon is well known since the seminal work by Risken et al. \cite{risken} who predicted that, for a fixed value of the friction coefficient $\gamma$, the locked and running solutions coexist in  the deterministic $\theta = 0$ counterpart of the system if the constant force is in the range
\begin{equation}
	f_1(\gamma) < f < f_3 = 1
\end{equation}
which is also indicated in Fig. \ref{fig5}. The force $f_1(\gamma)$ is a solution of the deterministic system as the minimal value $f$ for which a running state $v_r \neq 0$ starts to appear. Similarly, $f_2(\gamma)$ is determined as a force at which the mean velocity $\langle v \rangle$ jumps from zero to a value corresponding to a running solution $v_r$ in the limit of vanishing noise \mbox{$\theta \to 0$} \cite{risken}. On the line $f_2(\gamma)$ the locked $v_l = 0 $ and  running $v_r \neq 0$ states are equally probable in the deterministic limit of zero temperature. Above $f_2(\gamma)$ the running state is  more stable than the locked state and below 
$f_2(\gamma)$ the locked state is  more stable than the running state. The second class of non-monotonic temperature dependence of the diffusion coefficient which \emph{apparently} is not associated with the occurrence of the giant diffusion effect has been reported in Ref. \cite{lindner2016}, see Fig. 6 (b). In contrast to the divergence of the diffusion coefficient $D \to \infty$ with decreasing temperature $\theta \to 0$ in this second class the latter quantity goes to zero $D \to 0$ with temperature drop $\theta \to 0$. Therefore, \emph{apparently} the diffusion coefficient possesses distinct local maximum as a function of temperature.
\begin{figure}[t]
	\centering
	\includegraphics[width=0.9\linewidth]{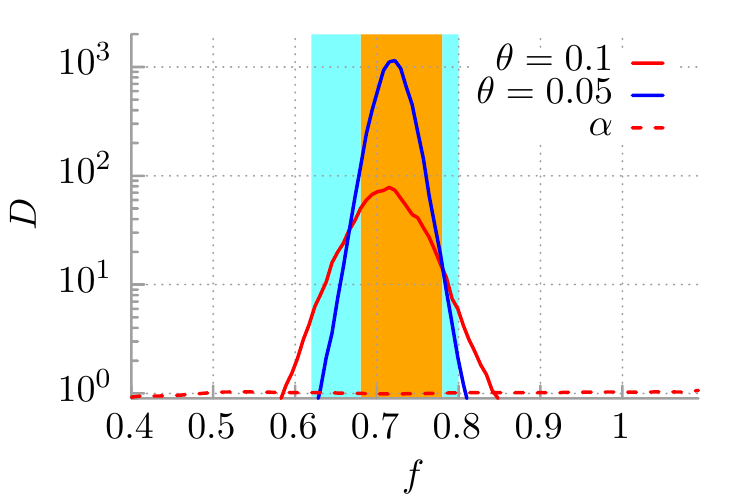}
	\caption{The diffusion coefficient $D$ versus the constant bias $f$ is plotted for $\gamma = 0.4$ and different values of temperature $\theta$. The power exponent $\alpha$ is plotted versus bias $f$ for $\theta = 0.1$. The colored region corresponds to the bias interval $f$ for which the diffusion coefficient $D$ displays non-monotonic character. The orange color indicates the area where according to Ref. \cite{lindner2016} the giant diffusion occurs.}
	\label{lindner}
\end{figure}
\begin{figure}[t]
	\centering
	\includegraphics[width=0.9\linewidth]{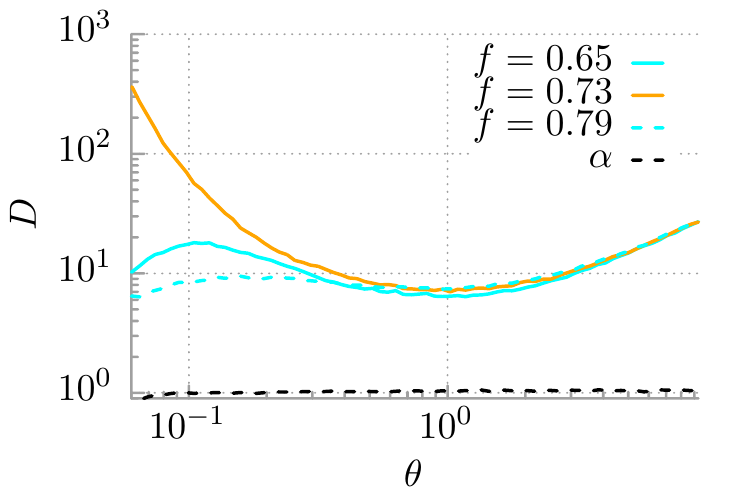}
	\caption{The diffusion coefficient $D$ versus temperature $\theta$ is shown for $\gamma = 0.4$ and several values of the bias $f$. The power exponent $\alpha$ is depicted for $f = 0.79$.}
	\label{lindner2}
\end{figure}

\subsection{Relation with the giant diffusion effect}
In \cite{lindner2016} the authors present the phase diagram for the occurrence of the enhancement of diffusion (sometimes called giant diffusion)
\begin{equation}
	D > D_0, \quad D_0 = \theta/\gamma. 
\end{equation}
in the limit of vanishing thermal noise intensity, see the orange region in Fig. 4 (b) of Ref. \cite{lindner2016}. The latter area is included in our diagram of the non-monotonic temperature dependence of diffusion, c.f. Fig. \ref{fig5}. However, it contains essentially new intervals where the non-monotonic behaviour is visible. To demonstrate this fact in Fig. \ref{lindner} we directly compare results depicted in Fig. 4 (a) of Ref. \cite{lindner2016} with the region of a non-monotonic temperature dependence of diffusion indicated in our diagram presented in Fig. \ref{fig5}. In the panel we plot the diffusion coefficient $D$ versus the constant bias $f$ for $\gamma = 0.4$ and different values of temperature $\theta$. In particular, the colored interval  of the force $f \approx [0.62,0.8]$ corresponds to the interval where the  diffusion coefficient $D$ becomes a non-monotonic function of temperature. The force interval 
$f \approx [0.68, 0.78] \subset [0.62,0.8]$ marked by orange color indicates the bias range for which the effect of giant diffusion occurs. It \emph{apparently} means that the diffusion coefficient diverges $D \to \infty$ as temperature goes to zero $\theta \to 0$. The cyan region corresponds to the second class of  non-monotonic behaviour in which diffusion \emph{apparently} goes to zero $D \to 0$ with temperature drop $\theta \to 0$ \cite{lindner2016}. In the phase diagram depicted in Fig. \ref{fig5} we do not distinguish these two scenarios as such classification would require low or extremely low temperatures which are not accessible numerically due to extremely long transient anomalous diffusion occurring in such regimes, c.f. Fig. \ref{fig4} (a). The latter classification may or may not be done depending on the duration of these transient effects even for a single parameter regime, not to mention the whole plane $(\gamma, f)$. Nevertheless the main conclusion is that the set $S_1$ of the parameter regimes for which giant diffusion occurs is contained in the set $S_2$ of the parameter values where the non-monotonic dependence is detected, i.e. $S_1 \subset S_2$. Additionally, in Fig. \ref{lindner2} we demonstrate the temperature dependence of the diffusion coefficient $D$ for three different values of the bias $f$ and $\gamma = 0.4$, c.f. Fig. \ref{lindner}. The magnitude $f = 0.65$ as well as $f = 0.79$ correspond to the cyan colored area in Fig. \ref{lindner} whereas $f = 0.73$ lies in the giant diffusion region marked with an orange color. In the considered parameter $\theta$ range for values below and above the giant diffusion threshold we observe for increasing temperature first a maximum and then a minimum in characteristic $D(\theta)$. In contrast, for $f = 0.73$ we note only a local minimum. For sufficiently high temperature $\theta \to \infty$ all curves overlap and are expected to recover the Sutherland-Einstein relation. The very low temperature limit $\theta \to 0$ is not presented here, but \emph{apparently} in the giant diffusion region $D$ is decreasing function of temperature whereas outside this region $D$ increases as temperature grows.

\section{Summary}
In this work we revisited the problem of underdamped Brownian motion in a biased periodic potential. We explain the mechanism responsible for the recently communicated peculiar diffusive behaviour when the diffusion coefficient displays non-monotonic dependence on the system temperature.

As the mechanism standing behind this counter-intuitive diffusive characteristic we identify temperature dependence of stationary probabilities for the particle to reside in the locked and running state both forming bistable velocity dynamics. In particular, overall spread of trajectories, i.e. also diffusion coefficient is large when for the biased potential there are many locked trajectories. The diffusion coefficient assumes its minimum when the stationary probability to reside in the locked state is minimal as well. However, we revealed a regime in the parameter space of the model exhibiting non-monotonic diffusive behaviour where velocity dynamics is multistable and possesses two running solutions and a locked one. This complexity of the noisy system has been overlooked in previous research on this topic. The role which is played by this additional running state in the temperature dependence of diffusion lies beyond the scope of this paper so we leave this problem open for future research.

Most importantly, our innovative computational method allowed us to construct a precise phase diagram for occurrence of the non-monotonic temperature dependence of diffusion in the parameter plane $(\gamma,f)$. We confronted the resulting plot with the phase diagram of the giant diffusion effect published in Ref. \cite{lindner2016}. It turned out that it significantly extends the previous prediction.

Last but not least, we highlight several controversies concerning the low temperature dependence of the diffusion coefficient. Due to extremely long transient anomalous diffusive behaviour they may never be numerically resolved at all and asymptotic analytical methods should be applied (which now are not yet elaborated). Current technical facilities allow to attack these problems at most for a single parameter regime and therefore we think that our diagram is still a significant step forward to complete understanding of underdamped diffusion in a biased washboard potential.

Summarizing, we have pointed out an unexpected property of the relatively simply, yet paradigmatic model of nonequilibrium statistical physics. The presented phase diagram may trigger experimental investigations aiming at corroboration of the non-monotonic temperature dependence of diffusion.

\section*{Acknowledgment}
This work has been supported by the Grant NCN No. 2017/26/D/ST2/00543 (J. S.)


\begin{thebibliography}{99}
	\bibitem{suther} W. Sutherland, Phil. Mag. 9, 781 (1905)
	\bibitem{ein} A. Einstein, Ann. Phys. 17, 549 (1905)
	\bibitem{hanggi2009} P. H\"anggi, F. Marchesoni, Rev. Mod. Phys. 81, 387 (2009)
	\bibitem{slapik2018} A. Slapik, J. {\L}uczka, J. Spiechowicz,  Commun. Nonlinear Sci. Numer. Simul. 55, 316 (2018)
	\bibitem{slapik2019} A. Slapik, J. {\L}uczka, P. H\"anggi, J. Spiechowicz, Phys. Rev. Lett. 122, 070602 (2019)
	\bibitem{slapik2019prappl} A. Slapik, J. {\L}uczka and J. Spiechowicz, Phys. Rev. Appl. 12, 054002 (2019)
	\bibitem{metzler2014} R. Metzler, J. H. Jeon, A. G. Cherstvy, E. Barkai, Phys. Chem. Chem. Phys. 16, 24128 (2014)
	\bibitem{petrov} Ju. N. Petrov, W. M. Fal'chenko, W. F. Mazanko, I. A. Yakubcov, S. P. Worona, Metallofizika 10, 124 (1988) 
	\bibitem{ganshin1999} A. N. Gan'shin \textit{et al.}, Low Temp. Phys. 25, 259 (1999)
	\bibitem{schuring2002} A. Sch{\"u}ring, S. M. Auerbach, S. Fritzsche and R. Haberlandt, J. Chem. Phys. 116, 10890 (2002)
	\bibitem{guo2014} M. Guo, H. Gelman and M. Gruebele, PLoS One 9, e113040 (2014)
	\bibitem{tung2016} W. S. Tung \textit{et al.}, ACS Macro Lett. 5, 735 (2016)
	\bibitem{iubini2015} S. Iubini, O. Boada, Y. Omar and F. Piazza, \textit{New J. Phys.} 17, 113030 (2015) 
	\bibitem{lee2015} Ch. K. Lee, J. Moix and J. Cao, J. Chem. Phys. 142, 164103 (2015)
	\bibitem{eltsov2006} V. B. Eltsov \textit{et al.}, Phys. Rev. Lett. 96, 215302 (2006)
	\bibitem{risken} H. Risken, \textit{The Fokker-Planck Equation: Methods of Solution and Applications} (Berlin-Heidelberg, Springer-Verlag, 1996)
	\bibitem{junction} A. Barone and G. Patern\`o, {\it Physics and Application of the Josephson Effect}, (Wiley, New York, 1982)
	\bibitem{Coffey} W.~T. Coffey, Yu. P. Kalmykov and J. T. Waldron, {\it The Langevin Equation}, 2nd edition, (World Scientific, Singapore, 2004) see Sects. 5 and 7-10 therein.
	\bibitem{Ful1975} P. Fulde, L. Pietronero, W.~R. Schneider, and S. Str\"assler, Phys.  Rev. Lett. 35, 1776 (1975); W. Dieterich, I. Peschel, and W.~R. Schneider, Z. Physik B 27, 177 (1977); T. Geisel, Sol. State Commun. 32, 739 (1979)
	\bibitem{Gru1981} G. Gr\"uner, A. Zawadowski, and P. M. Chaikin, Phys. Rev. Lett. 46, 511 (1981)
	\bibitem{denisov2014} S. Denisov, S. Flach and P. H\"anggi, Phys. Rep. 538, 77 (2014)
	\bibitem{constantini1999} G. Constantini and F. Marchesoni, Europhys. Lett. 48, 491 (1999)

	\bibitem{lindenberg2005} K. Lindenberg, A. M. Lacasta, J. M. Sancho and A. H. Romero, New. J. Phys. 7, 29 (2005)
	\bibitem{marchenko2014} I. G. Marchenko and I. I. Marchenko, Europhys. Lett. 100, 50005 (2012)
	\bibitem{kramer} J.C. Latorre,  G.A. Patriotism and  P.R. Kramer, J. Stat. Phys. 
	150, 776 (2013).
	\bibitem{marchenko2014a} I. G. Marchenko,  I. I. Marchenko, ans A. V. Ziglo, 
	Eur. Phys. J. B 87, 10 (2014)
\bibitem{lindner2016} B. Lindner and I. M. Sokolov, Phys. Rev. E 93, 042106 (2016)
	\bibitem{marchenko2017} I. G. Marchenko, I. I. Marchenko, V. I. Tkachenko, JETP Letters 106, 242 (2017); 109, 671 (2019)
	\bibitem{kubo1966} R. Kubo, Rep. Prog. Phys. 29, 255 (1966)
	\bibitem{spiechowicz2016scirep} J. Spiechowicz, J. {\L}uczka, P. H\"anggi, Sci. Rep. 6, 30948 (2016)	
\bibitem{spiechowicz2019njp} J. Spiechowicz, P. H\"anggi and J. {\L}uczka, New J. Phys. 21, 083029 (2019)
	\bibitem{dechant2019} A. Dechant, F. Kindermann, A. Widera and E. Lutz, Phys. Rev. Lett. 123, 070602 (2019)
	\bibitem{spiechowicz2019chaos} J. Spiechowicz and J. {\L}uczka, Chaos 29, 013105 (2019) 
	\bibitem{fisher2018} L. Fisher, P. Pietzonka and U. Seifert, Phys. Rev. E 97, 022143 (2018)
	\bibitem{cheng2018} Ch. Cheng, M. Cirillo, G. Salina and N. Gronbech-Jensen, Phys. Rev. E 98, 012140 (2018)
	\bibitem{kindermann2017} F. Kindermann, A. Dechant, M. Hohmann T. Lausch, D. Mayer, F. Schmidt, E. Lutz and A. Widera, Nat. Phys. 13, 137 (2017)
	\bibitem{spiechowicz2017scirep} J. Spiechowicz and J. {\L}uczka, Sci. Rep. 7, 16451 (2017)
	\bibitem{guerin2017} T. Guerin and D. S. Dean, Phys. Rev. E 95, 012109 (2017)
	\bibitem{zhang2017} J-M Zhang and J-D Bao, Phys. Rev. E 95, 032107 (2017)
	\bibitem{goychuk2019} I. Goychuk, Phys. Rev. Lett. 123, 180603 (2019)


	\bibitem{lindner2001} B. Lindner, M. Kostur and L. Schimansky-Geier, Fluct. Noise Lett. R25, 173 (2001)
	\bibitem{reimann2001a} P. Reimann, C. Van den Broeck, H. Linke, P. H\"anggi, J. M. Rubi and A. Perez-Madrid, Phys. Rev. Lett. 87, 010602 (2001)
	\bibitem{reimann2001b} P. Reimann, C. Van den Broeck, H. Linke, P. H\"anggi, J. M. Rubi and A. Perez-Madrid, Phys. Rev. E 65, 031104 (2001)


	\bibitem{spiechowicz2015pre} J. Spiechowicz and J. {\L}uczka, Phys. Rev. E 91, 062104 (2015)
	\bibitem{spiechowicz2016njp} J. Spiechowicz, P. Talkner, P. H\"anggi and J. {\L}uczka, New J. Phys. 18, 123029 (2016)
	\bibitem{spiechowicz2017chaos} J. Spiechowicz, M. Kostur and J. {\L}uczka, Chaos 27, 023111 (2017)
	\bibitem{marchenko2018} I. G. Marchenko, I. I. Marchenko and A. V. Zhiglo, Phys. Rev. E 97, 012121 (2018)
	\bibitem{spiechowicz2015cpc} J. Spiechowicz, M. Kostur and {\L}. Machura, Comp. Phys. Commun. 191, 140 (2015)
\end{thebibliography}
\end{document}